\documentclass[10pt,twocolumn,letterpaper]{article}

\usepackage[pagenumbers]{wacv} 

\usepackage{graphicx, float, wrapfig}
\usepackage{amsmath, amsthm, amssymb}
\usepackage{booktabs}
\usepackage{url} 
\usepackage{subcaption}                     
\usepackage[export]{adjustbox}              
\usepackage{multirow, enumitem}
\usepackage{xurl}                           
\usepackage{longtable, lscape}              
\usepackage{booktabs}                       
\usepackage{indentfirst}
\usepackage{xurl}

\usepackage{blindtext}
\newcommand\blfootnote[1]{%
\begingroup
\renewcommand\thefootnote{}\footnote{#1}%
\addtocounter{footnote}{-1}%
\endgroup
}

\usepackage[pagebackref,breaklinks,colorlinks]{hyperref}

\usepackage[capitalize]{cleveref}
\crefname{section}{Sec.}{Secs.}
\Crefname{section}{Section}{Sections}
\Crefname{table}{Table}{Tables}
\crefname{table}{Tab.}{Tabs.}

\def\wacvPaperID{2070} 
\def\confName{WACV}
\def\confYear{2025}

\begin{document}

\title{Coupling AI and Citizen Science in Creation of Enhanced Training Dataset for Medical Image Segmentation}

\author{Amir Syahmi*, Xiangrong L. Lu*, Yinxuan Li*, Haoxuan Yao*, Hanjun Jiang*,\\ 
Ishita Acharya, Shiyi Wang, Yang Nan, Xiaodan Xing, Guang Yang\textdagger\\[0.1cm]
Department of Bioengineering and Imperial-X, Imperial College London\\
London, W12 7SL, United Kingdom\\
\small \texttt{$\{$amir.syahmi21, xiangrong.lu21, yinxuan.li21, haoxuan.yao21, barry.jiang20,}\\
\small \texttt{ishita.acharya20, s.wang22, y.nan20, x.xing, g.yang$\}$@imperial.ac.uk}
}
\maketitle

\begin{abstract}
    Recent advancements in medical imaging and artificial intelligence (AI) have greatly enhanced diagnostic capabilities, but the development of effective deep learning (DL) models is still constrained by the lack of high-quality annotated datasets.
    The traditional manual annotation process by medical experts is time- and resource-intensive, limiting the scalability of these datasets.
    In this work, we introduce a robust and versatile framework that combines AI and crowdsourcing to improve both the quality and quantity of medical image datasets across different modalities.
    Our approach utilises a user-friendly online platform that enables a diverse group of crowd annotators to label medical images efficiently.
    By integrating the MedSAM segmentation AI with this platform, we accelerate the annotation process while maintaining expert-level quality through an algorithm that merges crowd-labelled images.
    Additionally, we employ pix2pixGAN, a generative AI model, to expand the training dataset with synthetic images that capture realistic morphological features.
    These methods are combined into a cohesive framework designed to produce an enhanced dataset, which can serve as a universal pre-processing pipeline to boost the training of any medical deep learning segmentation model.
    Our results demonstrate that this framework significantly improves model performance, especially when training data is limited.
\end{abstract}

\section{Introduction}
\blfootnote{*Contributed equally}
\blfootnote{\textdagger Correspondence Author}
\blfootnote{The code for this proposed framework is publicly available on GitHub: \url{https://github.com/Am1rSy/SGC-Enhanced-Dataset}.}
\label{sec:introduction}
\begin{figure*}
    \centering
    \includegraphics[width=0.85\linewidth]{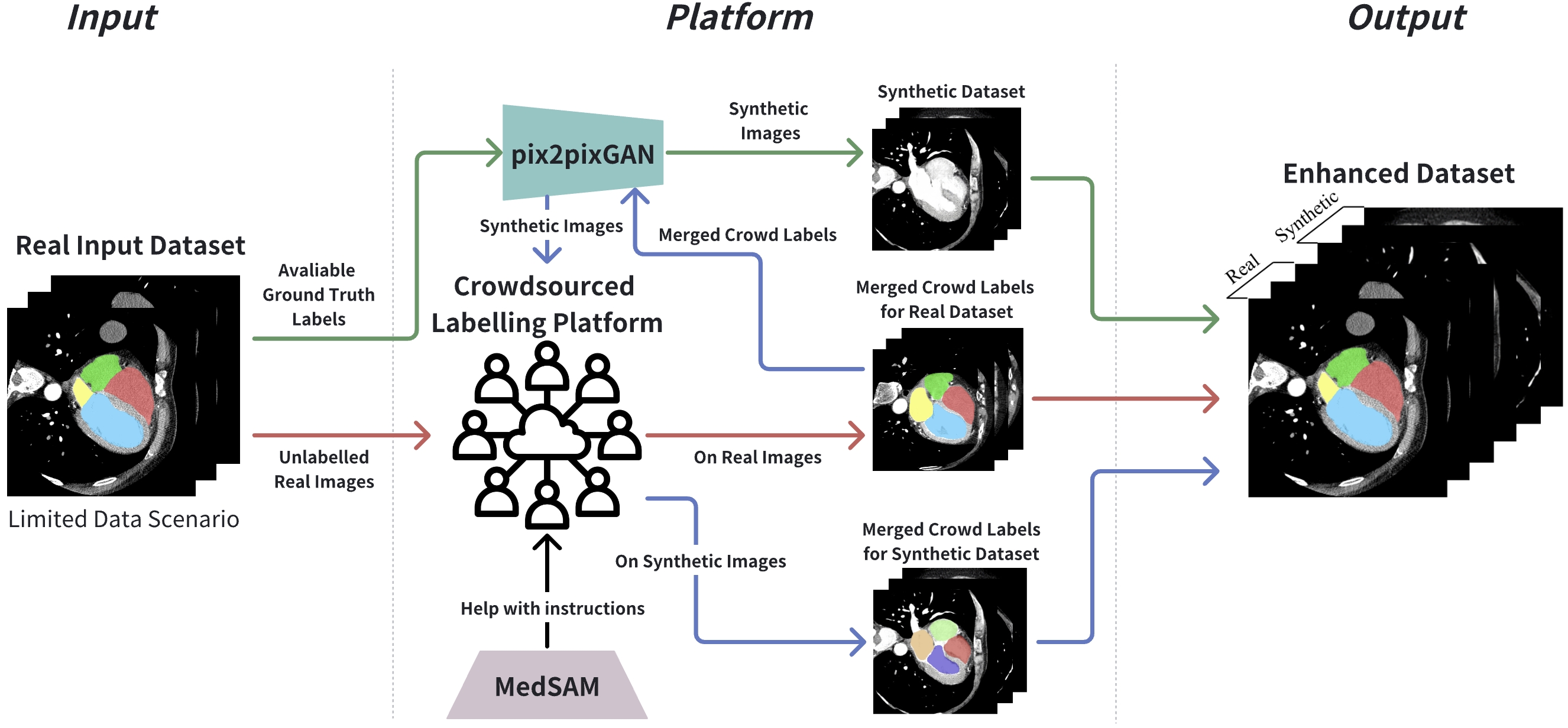}
    \caption{Summary of the proposed framework: Real medical image datasets are provided to a crowdsourced labelling platform and also used to train pix2pixGAN. Annotators would label the real dataset images with assistance from MedSAM. These annotated real images are then used by pix2pixGAN to generate synthetic images, which are subsequently sent back to the labelling platform for further annotation. The labelled synthetic and real images are finally combined into an enhanced dataset. Figure illustrated by Lark}
    \label{meth:Framework}
\end{figure*}
The rapid advancements in medical imaging technologies, such as CT and MRI, have revolutionised diagnostic capabilities in healthcare by enabling non-invasive visualisation of internal anatomical structures. 
Alongside this, the recent progress in artificial intelligence (AI) has led to growing interest and significant improvements in developing medical image analysis algorithms, which subsequently improve automated diagnostics. 
These advancements enable healthcare professionals to make more precise diagnoses and develop more effective treatment plans\cite{heimann2009statistical}\cite{zhuang2019evaluation}\cite{walsh2020imaging}.

Image segmentation is one of the most commonly used analysis techniques and involves the delineation of structures and forms the building blocks of any AI-assisted image labelling algorithm. 
However, a major limiting factor to the development of a robust and accurate AI-based medical image analysis model is the lack of high-volume and high-quality annotated training datasets\cite{alzubaidi2023survey}. 

High-quality labelled datasets are crucial as they directly influence the accuracy and reliability of AI models, ensuring these technologies can make precise diagnostic predictions\cite{galbusera2024image}.
Generating these labelled datasets is both time-consuming and resource-intensive, making scalability a challenge\cite{woznicki2023addressing}. 
Furthermore, hiring medical experts to manually label a large quantity of medical images is costly, and the process is often tedious due to its repetitive nature\cite{woznicki2023addressing}. 
Thus, crowdsourcing has been seen as an attractive method to help improve the annotation rate for medical images. 
It operates based on allowing untrained individuals to help annotate new, unlabelled datasets\cite{raykar2010learning}. 
Researchers have explored the potential of crowdsourcing to cost-effectively expand the annotated datasets for medical images\cite{petrovic2020crowdsourcing}. 
Studies have shown that with clear instructions and examples, non-experts can achieve performance that matches that of trained experts, particularly for certain imaging modalities\cite{gurari2015collect}. 
However, the complexity of the medical images still needs to be investigated to understand the limitations of crowd-sourcing fully. 

The challenge of acquiring a substantial volume of medical images for training AI solutions is another significant bottleneck in the field. 
This is due to the high cost and logistical complexity involved in producing such datasets, which require specialised equipment and trained personnel. 
Additionally, privacy concerns are also limiting the acquisition due to the handling of sensitive personal information, which usually requires extra data masking procedures. 
The diversity of medical imaging modalities (e.g. CT, MRI) further complicates the collection process, as acquiring comprehensive data across all these modalities is a daunting task\cite{greenspan2016guest}. 
Although, the release of datasets publicly by various research groups in recent years (MMWHS\cite{zhuang2022MMWHS}, FLARE22\cite{ma2022FLARE22}, Cancer Imaging Archive\cite{NCICIP2022Cancer}) has somewhat mitigated the issue, given the data-intensive nature of AI model training, particularly with Deep Learning (DL) approaches, the demand for extensive datasets remains unabated\cite{sarker2021deep}. 
Consequently, exploring the potential of generative AI to augment real datasets by creating synthetic but close-to-realistic images presents a promising area of research\cite{musalamadugu2023generative}. 
This approach can help overcome the inherent limitations of insufficient variety and volume of real datasets, by generating diverse and extensive training data. 
Such synthetic data can improve the training of AI models, enabling them to help achieve higher accuracy and generalisability in medical image analysis\cite{ali2022role}.

In this work, we present a versatile framework that enhances medical image dataset in both quality and quantity by coupling crowdsourcing and generative AI.
This method ultimately increases segmentation accuracy of DL models when available training dataset is limited.

\section{Related Works}
Crowdsourcing in image analysis, seen in Google's reCAPTCHA\cite{Google2024reCAPTCHA} and Duolingo\cite{Duolingo2024Fraud}, also applies to biomedical imaging. 
Studies show that crowds can accurately label complex medical images, demonstrating the potential for medical image analysis. 
Platforms like Amazon Mechanical Turk (MTurk)\cite{Amazon2024AWSMTurk} and Appen Figure Eight\cite{Appen2024Fig8} streamline crowdsourcing by providing a diverse pool of participants and reducing the need for custom setups. 
Alternatively, custom platforms like Label Studio\cite{HumanSignal2024LabelStudio}, though more time-intensive to develop, offer precise control over the crowdsourcing tasks, improving the engagement and specificity of the work. 
In summary, crowdsourcing in image analysis extends its utility to biomedical fields, demonstrating significant potential in medical diagnostics. 
By utilising diverse participant pools and flexible setup options, these methods elevate data accuracy and streamline the labelling process, and prove essential for advancing accurate DL approach for medical image segmentation.

Moreover, there has been a lot of research in using generative AI to improve data augmentations\cite{xing2024aieatsitselfcaveats}\cite{Shorten2019}.
Specifically, Generative Adversarial Network (GAN)-based models\cite{goodfellow2014generative}\cite{goodfellow2020generative} are widely used for synthesising different medical images, successfully expanding the size of biomedical imaging datasets\cite{creswell2018generative}. 
Deep Convolutional GAN (DCGAN), an unconditional GAN-based model, has been used to generate high-quality liver lesion region of interests (ROIs)\cite{frid2018gan}. 
These synthetic images, combined with real images, are used to train Convolutional Neural Networks (CNNs) and greatly improve the performance of CNN classification models, thereby enhancing diagnosis. 
Additionally, synthetic MRI images generated by a conditional GAN model have successfully boosted the training of segmentation networks and improved the performance of brain tumour MR image segmentation\cite{mok2019learning}.

\begin{figure*}
    \centering
    \includegraphics[width=0.85\linewidth]{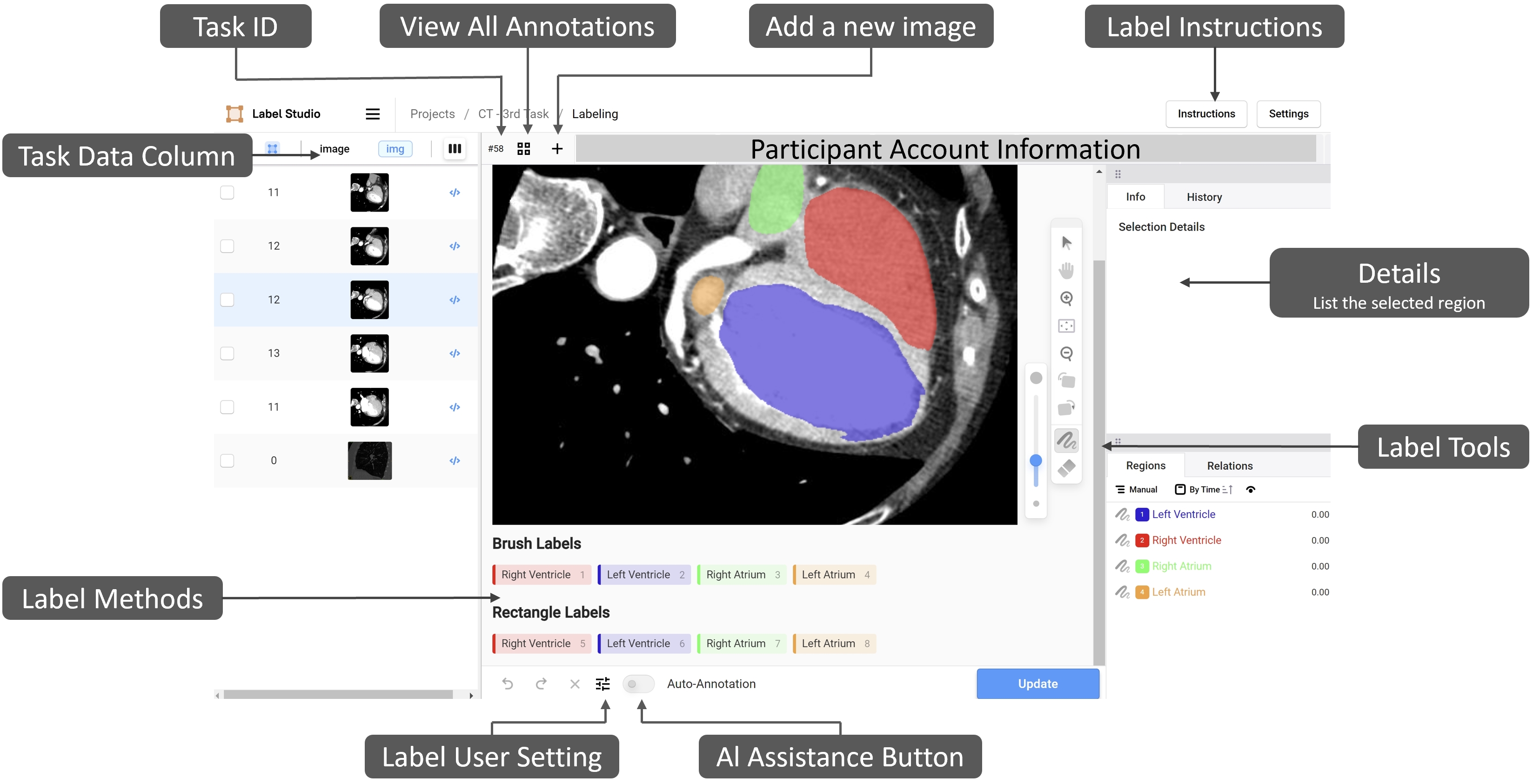}
    \caption{Label Studio UI}
    \label{meth:LabelStudioUI}
\end{figure*}

\section{Our Approach}
\label{sec:methodologies}
\subsection{Main Contributions}

By coupling AI and citizen science, we aim to improve the data gathering rate and create an extensive and labelled dataset for future medical image DL research.
This enhancement can be achieved by: deploying a flexible segmentation model to facilitate the labelling process, thereby reducing both time and tedium for crowd annotators; enlisting the aid of the public via crowdsourcing, with adequate guidance, to accelerate the annotation rate; implementing generative AI to expand the medical image datasets with synthetic images.

By achieving the above aims, we highlight 4 key contributions of this work:
\begin{enumerate}[noitemsep,nolistsep]
    \item We proposed a \emph{versatile framework} to efficiently and effectively resolve the scarcity and costliness of medical image datasets for DL model training.
    \item We implemented a state-of-the-art \emph{segmentation AI} MedSAM to simplify crowdsourced segmentation which attains labelling at an expert \emph{quality}.
    \item We incorporated a novel \emph{generative AI} pix2pixGAN to expand the \emph{quantity} of existing dataset in different modalities.
    \item We verified that, using the framework we proposed, the \emph{accuracy} and \emph{performance} of DL models for medical images would be significantly improved with \emph{small} training dataset. 
\end{enumerate}

In short, we established a \textbf{versatile} framework that can expand limited medical image dataset in \textbf{quantity} and also with similar label \textbf{quality} as the domain experts.
Such dataset will be called \textbf{enhanced dataset} (quality + quantity) for the rest of this work.
An overview of our framework can be seen in Figure \ref{meth:Framework}.

\subsection{Crowd Labelling Platform Deployment}
Image labelling tasks for crowds, especially with medical images, are often complex and result in a lack of accuracy and willingness to participate. 
We first needed to implement an online platform for the ease of communication and labelling operation from the researchers to a wide audience with various types of device. 
Label Studio was chosen as the main data labelling platform as it is open source and contains a simple, friendly user interface (UI) (See Figure \ref{meth:LabelStudioUI})\cite{HumanSignal2024LabelStudio}. 
An easily navigable UI is key in this study as the labelling process needs to be straightforward to account for various computer literacy in the public.
We designed the platform to allow the use of a few tools which includes labelling brush, eraser, zooming, etc.
Furthermore, we provided all users with an instructional PDF file containing ground truth label exemplars and a short video to guide them on using the platform (see Supplementary Section 3 and 4).
Label Studio was also chosen because it is easily deployed on online servers\cite{HumanSignal2024LabelStudio}, a feature well supported by its extensive developer community. 
Our current implementation of Label Studio is hosted on the Hugging Face Spaces platform\cite{HugginFaceE2024HugginFaceSpaces}. 
This hosting platform was selected for its capability to support application deployment using Docker containers.

\subsection{Segmentation AI Integration}
Segmentation AI assistance has been proven to be an effective approach to further resolve the complexity of labelling\cite{heim2018large}. 
However, the use of segmentation AI aiding in the labelling process needs to be versatile for a wide range of tasks regarding medical images, and intuitive for the users to operate. 
Label Studio\cite{HumanSignal2024LabelStudio} also supports AI integration during the labelling process. 
It operates by using another server as a Machine Learning (ML) backend, where the AI model is hosted. 
We chose MedSAM\cite{ma2024segment} for our AI integration (see Section \ref{result:Quality:Segmentation AI Comparison for Crowd Use}). 
MedSAM is a zero-shot segmentation model based on Meta's Segment Anything model (SAM)\cite{kirillov2023segment}  that has been trained on 1 million medical images.
For integration with the labelling platform, we only allow the bounding box functionality to appear when a toggle switch is activated. 
A user would select the rectangular label from the available selection and draw a box on the image (see Supplementary Section 3).
Then, Label Studio will send the necessary information (bounding box coordinates and current image) to MedSAM. 
MedSAM will consider the spatial relationship and contextual information inherent in medical images when processing the information for segmentation\cite{ma2024segment}.
Finally, it will send its predicted labels of the specified region back to Label Studio. 
This would allow for faster and more accurate labelling created by the users.

\subsection{Generative AI Integration}
As crowd-labelling methods are limited by the availability of raw medical images, generative AI, particularly Generative Adversarial Networks (GANs)\cite{goodfellow2014generative}\cite{goodfellow2020generative}, is used for data augmentation purposes.
Using the GAN model of our choice, synthetic medical images are generated using labels provided by the crowd and from the input dataset.
To achieve this, the GAN model can be extended to a conditional GAN (cGAN) model\cite{mirza2014conditional}.
A condition will be given to both the generator and the discriminator during the training process to partially control the generated images.
We used user-generated labels from crowdsourcing as input into the cGAN.
As the labels are in image format, a variant of cGAN named pix2pixGAN\cite{isola2017image} was adapted for our project.
This is due to pix2pixGAN being specially designed to solve image-to-image translation problems, where the input consists of image-label pairs.
These synthetic images generated by pix2pixGAN are then integrated with the annotated image dataset to meet the needs of training future medical Image models.

\subsection{Test Trials}
The framework was tested on 3 different datasets which are MMWHS-MRI, MMWHS-CT and FLARE22\cite{zhuang2022MMWHS}\cite{ma2022FLARE22}. 
We recruited 12 annotators to investigate the effectiveness of the general public in labelling medical images, and their results were further used to verify the potential of improving DL model training with crowdsourcing. 
We assigned each annotator 6 tasks, containing 5 images each. The objective and criteria of each task are listed as follows:
\begin{enumerate}[noitemsep,nolistsep]
    \item Task 1: Labelling of the specified heart regions (MMWHS-CT) without any AI assistance or ground truth exemplars.
    \item Task 2: Labelling of the specified heart regions (MMWHS-CT) with AI assistance but no ground truth exemplars are provided.
    \item Task 3: Labelling of the specified heart regions (MMWHS-CT) with AI assistance and ground truth exemplars.
    \item Task 4: Labelling of the specified heart regions of the artificial GAN-generated dataset with AI assistance.
    \item Task 5: Labelling of the specified abdominal organs (FLARE22) with AI assistance and ground truth exemplars.
    \item Task 6: Labelling of the specified heart regions (MMWHS-MRI) with AI assistance and ground truth exemplars.
\end{enumerate}
Tasks 1, 2, and 3 serve to evaluate the necessity of AI assistance and instructions with ground truth exemplars in crowdsourcing platforms. 
Task 4 serves to assess the crowdsourcing label performance on GAN generated dataset. 
Tasks 3, 5, and 6 serve to understand the versatility of the platform in various datasets.
(Detailed results see Supplementary Section 6)

\subsection{Merged Crowd Labels}
\label{meth:Merged Crowd Annotations}
To combine the ensemble of crowd labels of a single image into a merged label, a pixel-wise majority voting approach is taken\cite{maier2014can}. 
In this approach, the labels of each pixel are summed to create a frequency map. 
Subsequently, a threshold is applied to this map to generate the merged label. 
This threshold represents the minimum number of crowd annotators required to agree that a specific pixel belongs to the object of interest. 
In this work, a minimum threshold of 4 was chosen based on the consideration of the number of unique crowd annotators.

\subsection{Evaluation Metrics - Comparison with Ground Truth}
To evaluate the quality of the segmentations results, the Sørensen-Dice index (DSC)\cite{sorensen1948method} and the Jaccard Index (IoU)\cite{sorensen1948method} are commonly used.
These metrics, defined in Equation \ref{meth:eqn:DSC} and \ref{meth:eqn:IoU} respectively, are selected due to their widespread usage and ease of comparison with other publications and methods.

\vspace{-0.5cm}
\begin{minipage}[t]{0.49\linewidth}
        \begin{equation}
            D(X,Y) = \frac{2\left|X \cap Y \right|}{\left|X  \right| + \left|Y \right|}
            \label{meth:eqn:DSC}
        \end{equation}
\end{minipage}
\begin{minipage}[t]{0.45\linewidth}
    \begin{equation}
        J(X,Y) = \frac{\left|X \cap Y \right|}{\left|X \cup Y \right|}
        \label{meth:eqn:IoU}
    \end{equation}
\end{minipage}

\section{Experiment}
\label{sec:result}
The experiment section is comprised of 3 sub-sections.
Section \ref{result:Quality} contains our initial findings on the framework and evaluates the performance of crowd labellers in achieving expert-level labelling.
Section \ref{result:Quantity} evaluates the performance of an enlarged dataset using synthetic images generated by pix2pixGAN.
Lastly, Section \ref{result:temp} evaluates the effectiveness of training DL segmentation models by combining crowd-labelled images and synthetic images into one enhanced dataset, which is the overall outcome of the framework.

\begin{figure*}
    \centering
    \includegraphics[width=0.75\linewidth]{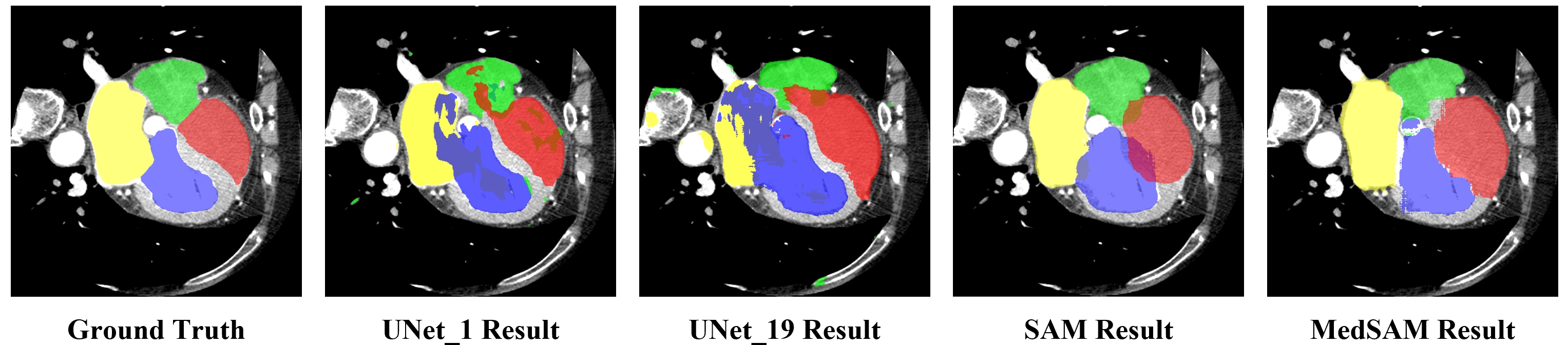}
    \caption{
        Comparison of the results for, from Left to Right, Ground Truth, UNet\_1, UNet\_19, SAM, MedSAM on MMWHS-CT slice 110
    }
    \label{res:SegmentationComparison}
\end{figure*}

\begin{table*}
    \centering
    \scalebox{0.7}{
        \begin{tabular}{ccccccccc}
            \toprule[1.2pt]
            & \multicolumn{2}{c}{UNet\_1} & \multicolumn{2}{c}{UNet\_19} & \multicolumn{2}{c}{SAM} & \multicolumn{2}{c}{MedSAM} \\
            ROI     & DSC           & IoU          & DSC           & IoU         & DSC           & IoU          & DSC           & IoU           \\
            \midrule[1.2pt]
            LA      & 0.5268        & 0.5232       & 0.4535        & 0.4721      & 0.6079     & 0.5149     & 0.7129       & 0.6306      \\
            RA      & 0.4109        & 0.4970       & 0.2986        & 0.3522      & 0.7253     & 0.6102     & 0.7827       & 0.6860      \\
            LV      & 0.7749        & 0.8909       & 0.7973        & 0.7675      & 0.7926     & 0.6639     & 0.8929       & 0.8070      \\
            RV      & 0.8509        & 0.8943       & 0.6574        & 0.6840      & 0.8118     & 0.6840     & 0.8605       & 0.7555      \\\hline
            Average & 0.6409        & 0.7014       & 0.5517        & 0.5690      & 0.7344     & 0.6183     & 0.8123       & 0.7197      \\
            \bottomrule[1.2pt]
        \end{tabular}
    }
    \caption{Mean DSC and IoU result for, from Left to Right, UNet\_1, UNet\_19, SAM, and MedSAM}
    \label{tab:UNetComparison}
\end{table*}

\subsection{Quality -- Achieving Image Crowd Labelling at a Professional Level}
\label{result:Quality}

\subsubsection{Segmentation AI Comparison for Crowd Use}
\label{result:Quality:Segmentation AI Comparison for Crowd Use}

The primary objective of incorporating an AI assistance labelling tool on our platform is to improve the efficiency and ease of the segmentation task for crowd annotators.
As a preliminary study, we investigated the most suitable segmentation AI model that is capable of assisting users in labelling tasks\footnote{Abbreviation of the ROIs: LA - Left Atrium; RA - Right Atrium; LV - Left Ventricle; RV - Right Ventricle}.
A comparative analysis is conducted on prediction masks generated by 4 segmentation models: UNet\_1, UNet\_19, SAM, and MedSAM.
Both UNet\_1 and UNet\_19 are based on simple UNet structure, UNet\_1 is trained on 10 training slices from 1 sub-dataset of MMWHS-CT, whereas UNet\_19 is trained on 76 training slices from 19 sub-datasets of MMWHS-CT.

Figure \ref{res:SegmentationComparison} shows example prediction masks generated by UNet\_1 and UNet\_19 models on MMWHS-CT slice 110.
Notably, both predicted masks are characterized by un-smooth and irregular contours with small, scattered regions due to overlapping labels. 
These graphical observations are confirmed by the metrics in Table \ref{tab:UNetComparison}, where UNet\_1 outperforms UNet\_19 with a higher overall metric score. 
Both models achieve relatively high metric scores above 0.65 for ventricle labels and relatively low scores below 0.53 for atrium labels. 
Results from Figure \ref{res:SegmentationComparison} and Table \ref{tab:UNetComparison} suggest that the UNet models are unsuitable for platform AI assistance due to their poor versatility across different datasets. 
Each label task in the platform requires a new UNet model specifically trained for the corresponding dataset, and even ideally, the sub-datasets, despite the same modality and similar morphological structure. 
This is validated by the superior performance of UNet\_1 over UNet\_19, which was achieved even with less training data and reduced training time. 

SAM and MedSAM are tested as large-scale models without specific training on any MMWHS datasets.
Figure \ref{res:SegmentationComparison} also illustrates the prediction masks of the same testing slice generated by SAM and MedSAM models, characterized by smooth contours and significantly fewer overlapping regions.
These observations are corroborated by metric scores detailed in Table \ref{tab:UNetComparison}. 
In contrast to the UNet models, which demonstrated higher performance of ventricle labels over atrium labels, both SAM and MedSAM exhibit consistent performance across all labels, which demonstrate their high versatility.
Specifically, SAM achieves an average DSC of 0.7344 (IoU of 0.6183), while MedSAM reaches an average DSC of 0.8123 (IoU of 0.7197). 
MedSAM outperforms SAM and the UNet models in graphical representation and metric evaluations and hence, the chosen segmentation AI for crowd-labelling.

\begin{table*}
    \vspace{-0.1cm}
    \centering
    \scalebox{0.7}{
        \begin{tabular}{ccccccc}
            \toprule[1.2pt]
            & \multicolumn{2}{c}{Task 1}   & \multicolumn{2}{c}{Task 2}    & \multicolumn{2}{c}{Task 3} \\
            ROI     & DSC           & IoU          & DSC           & IoU      & DSC           & IoU     \\
            \midrule[1.2pt]
            \multirow{2}{*}{LA} & 0.3106              & 0.2874              & 0.3113                 & 0.2842             &0.4272       &0.4003                    \\
                        & {[}0.1876 0.4336{]} & {[}0.1726 0.4021{]} & {[}0.1950 0.4276{]}    & {[}0.1760 0.3924{]}    &{[}0.3002 0.5542{]}  &{[}0.2793 0.5213{]}   \\
            \multirow{2}{*}{RA} & 0.4132              & 0.3511              & 0.5093                 & 0.4327             &0.7020       &0.6161                    \\
                        & {[}0.3014 0.5250{]} & {[}0.2528 0.4494{]} & {[}0.4019 0.6167{]}    & {[}0.3359 0.5295{]}    &{[}0.6143 0.7898{]}  &{[}0.5340 0.6983{]}   \\
            \multirow{2}{*}{LV} & 0.6661              & 0.6010              & 0.5332                 & 0.4521             &0.8722       &0.7871                    \\
                        & {[}0.5740 0.7581{]} & {[}0.5139 0.6880{]} & {[}0.4391 0.6272{]}    & {[}0.3677 0.5364{]}    &{[}0.8460 0.8984{]}  &{[}0.7482 0.8261{]}   \\
            \multirow{2}{*}{RV} & 0.6513              & 0.5729              & 0.6506                 & 0.5924            &0.8877        &0.8112                    \\
                        & {[}0.5624 0.7402{]} & {[}0.4926 0.6533{]} & {[}0.5512 0.7499{]}    & {[}0.5001 0.6847{]}    &{[}0.8569 0.9185{]}  &{[}0.7793 0.8431{]}   \\
            \bottomrule[1.2pt]
        \end{tabular}
    }
    \vspace{-0.1cm}
    \caption{Mean DSC and IoU with 95\% CI for results collected from Task 1 (Hand-drawn), Task 2 (with AI assistance), and Task 3 (with AI assistance and instructions)}
    \label{tab:T1vT2prism}
\end{table*}

\begin{figure*}
    \vspace{-0.7cm}
    \begin{minipage}[b]{0.3\linewidth}
        \begin{table}[H]
            \centering
            \scalebox{0.7}{
                \begin{tabular}{lcc}
                    \toprule[1.2pt]
                    ROI     & DSC           & IoU           \\
                    \midrule[1.2pt]
                    LA & 0.3839 & 0.3627 \\
                    RA & 0.8504 & 0.7531 \\
                    LV & 0.7246 & 0.6622 \\
                    RV & 0.9159 & 0.8454 \\
                    \bottomrule[1.2pt]
                \end{tabular}
            }
            \caption{Mean DSC and IoU for merged labels from Task 3 MMWHS-CT}
            \label{tab:T2vT3prism_merged}
        \end{table}
    \end{minipage}
    \hspace{0.025\linewidth}
    \begin{minipage}[b]{0.3\linewidth}
        \begin{table}[H]
            \centering
            \scalebox{0.7}{
                \begin{tabular}{lcc}
                    \toprule[1.2pt]
                    ROI              & DSC    & IoU    \\
                    \midrule[1.2pt]
                    Liver  & 0.9698 & 0.9415 \\
                    Kidney & 0.7597 & 0.6536 \\
                    Aorta  & 0.9584 & 0.9204 \\
                    \bottomrule[1.2pt]
                \end{tabular}
            }
            \caption{Mean DSC and IoU for merged label from Task 5 FLARE22 Abdomen}
            \label{tab:T5FLARE22}
        \end{table}
    \end{minipage}
    \hspace{0.025\linewidth}
    \begin{minipage}[b]{0.3\linewidth}
        \begin{table}[H]
            \centering
            \scalebox{0.7}{
                \begin{tabular}{lcc}
                    \toprule[1.2pt]
                    ROI              & DSC    & IoU    \\
                    \midrule[1.2pt]
                    LV               & 0.6966 & 0.5850 \\
                    Myocardium of LV & 0.7197 & 0.6260 \\
                    RV               & 0.7146 & 0.6472 \\
                    Pulmonary Artery & 0.7382 & 0.5917 \\
                    \bottomrule[1.2pt]
                \end{tabular}
            }
            \caption{Mean DSC and IoU for merged label from Task 6 MMWHS-MRI}
            \label{tab:T6CT}
        \end{table}
    \end{minipage}
    \hfill
\end{figure*}

\subsubsection{Necessity of AI Assistance and Instruction}
\label{result:Quality:Necessity of AI Assistance and Instruction}
To investigate if AI assistance can improve crowd segmentation results, the comparison and analysis between the results from Task 1 and Task 2 (Hand-drawn vs. AI assistance) are as follows, using the DSC and IoU metrics. 

\vspace{-0.3cm}
\begin{figure}[H]
    \centering
    \includegraphics[width=0.7\linewidth]{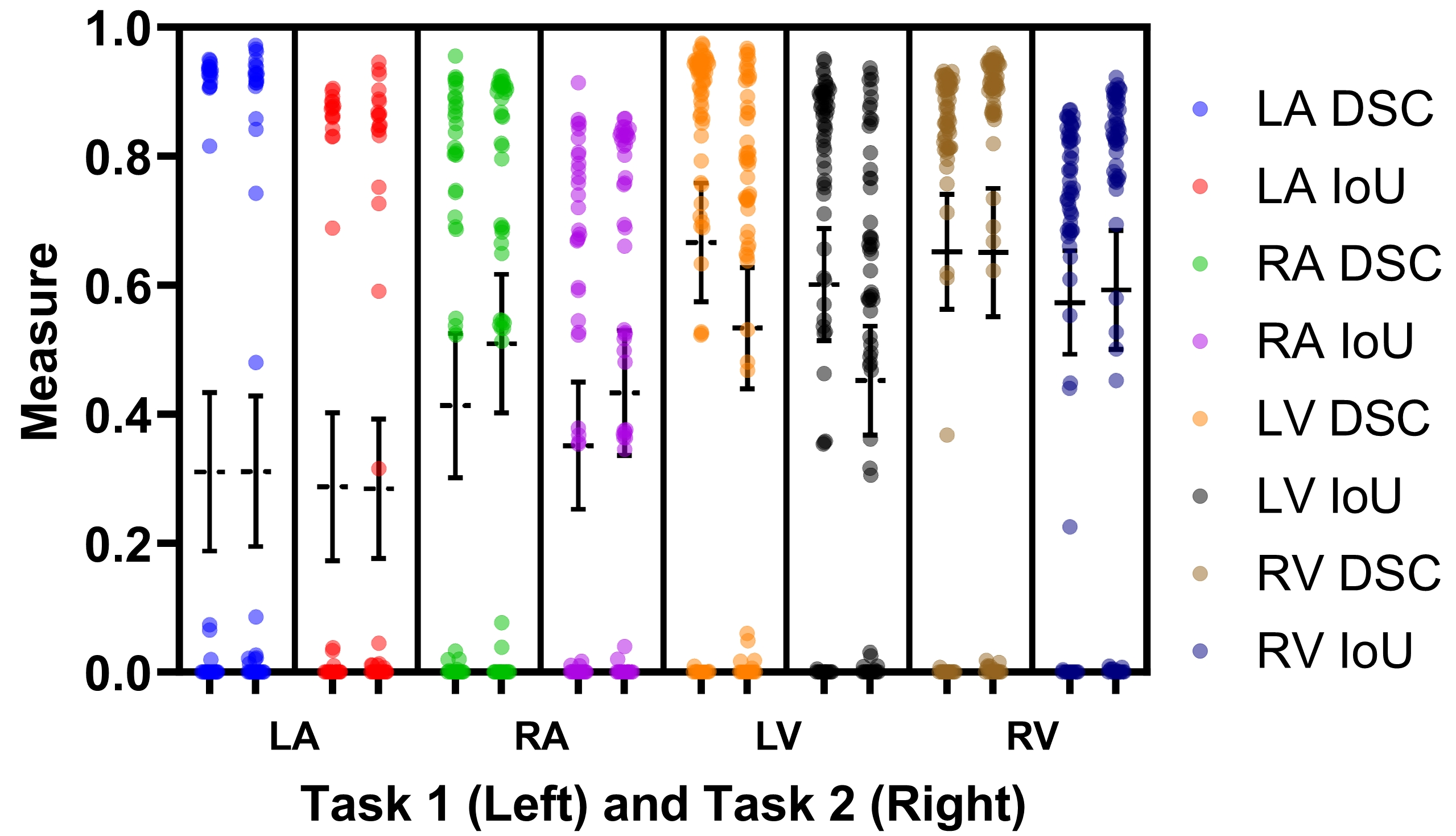}
    \caption{
        Comparison of DICE and IoU between each ROI for Task 1 (hand-drawn) and Task 2 (AI assistance); illustrated by Prism GraphPad 
    }
    \label{res:T1vT2prism}
\end{figure}
\vspace{-0.5cm}

From Figure \ref{res:T1vT2prism}, it is demonstrated that the distribution of metrics scores does not vary significantly between Task 1 and Task 2.
A noticeable amount of data points clustered around 0 is observed.
Quantitatively in Table \ref{tab:T1vT2prism}, it is statistically evident that, for all compartments of the heart, crowd segmentation accuracy from Task 1 and Task 2 are not significantly different ($p>0.05$).
This indicates that with only segmentation AI assistance provided, the accuracy of the crowd result would not be improved.
It is hypothesised that most of the volunteers have no prior knowledge of heart anatomy, leading to almost random annotations that do not fit with the ground truth.
Furthermore, some users also reported being confused with the orientation of the images.
It should be noted that once participants became accustomed to the AI tool, the majority of users reported an easier labelling process and reduction in labelling time by simply making quick refinements on the AI-generated regions.
This result highlights the success of making the segmentation process easier and less tedium with the deployment of the MedSAM segmentation facilitated tool.
To seek for actual improvement in accuracy, we hypothesised that an instruction, in addition to AI assistance, would provide fundamental knowledge to users that will in turn increase labelling accuracy.

To investigate whether the crowd segmentation results would improve when crowd annotators receive AI assistance along with detailed instructions including ground truth label exemplars.
The results from Task 2 and Task 3 (AI assistance vs. AI assistance with instructions) using DSC and IoU metrics are computed as follows.
\vspace{-0.3cm}
\begin{figure}[H]
    \centering
    \includegraphics[width=0.7\linewidth]{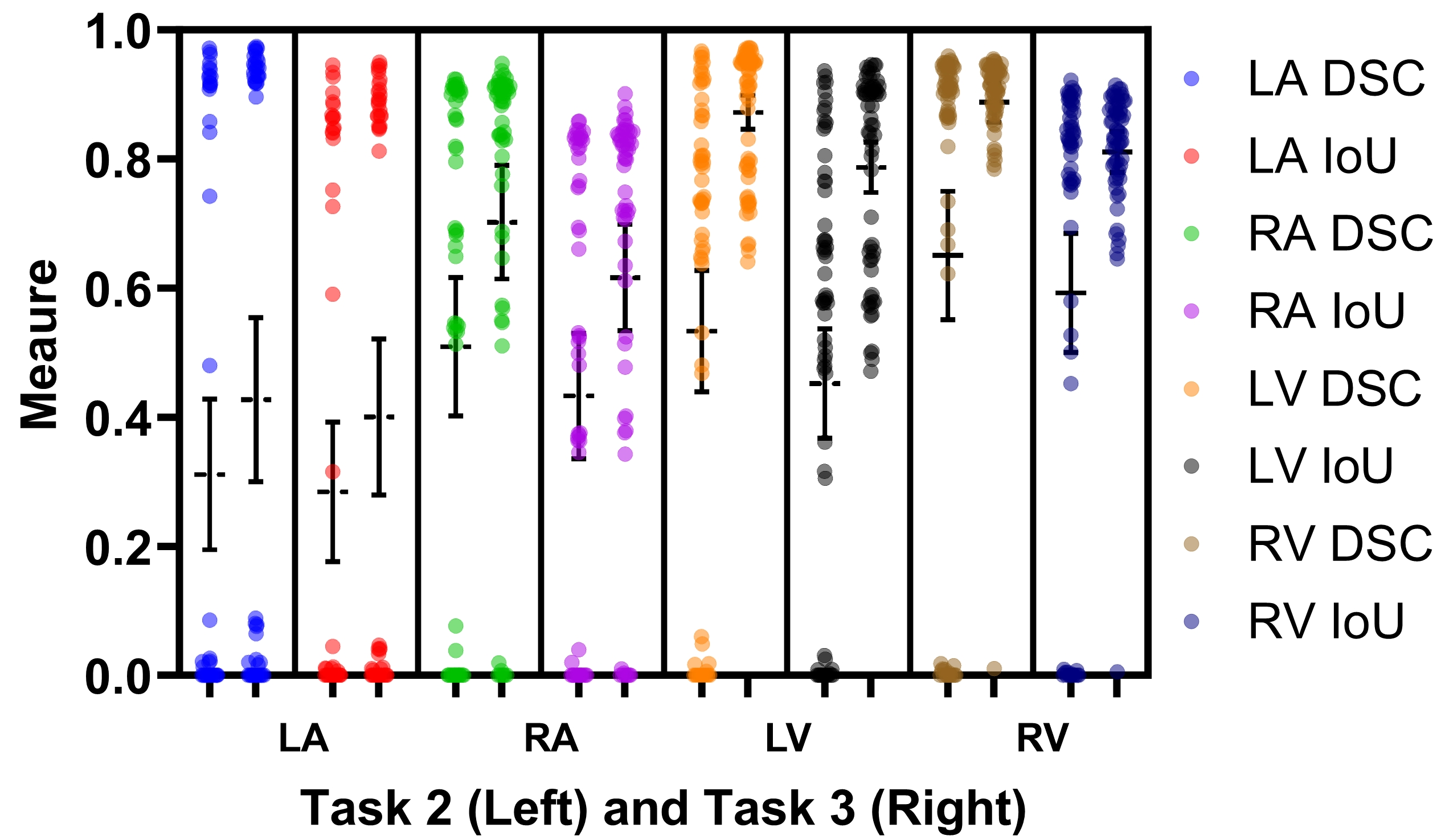}
    \caption{
        Comparison of DICE and IoU between each ROI for Task 2 (AI-assisted) and Task 3 (AI-assisted and instructions with ground truth exemplars); illustrated by Prism GraphPad 
    }
    \label{res:T2vT3prism}
\end{figure}
\vspace{-0.5cm}

It is evident in Table \ref{tab:UNetComparison} that the crowd segmentation results from Task 3 are statistically more accurate than those from Task 2 with 95\% CI ($p<0.05$), demonstrating that the instructions with ground truth exemplars are crucial to the accuracy of crowd labelling outcomes.
To account for the variance between annotators, we merged the crowd labels using pixel-wise majority voting approach with threshold of 4 (as discussed in Section \ref{meth:Merged Crowd Annotations}).
Table \ref{tab:T2vT3prism_merged} shows the metrics after merging.
Notably, LA has a relatively low DSC of 0.3839 (IoU of 0.3627), indicating the difficulty in labelling this ROI.
Hence, it is demonstrated that crowd annotators tend to perform better with simple anatomical structures that have less variance between slices.
This observation suggests that crowdsourcing should be limited to datasets with relatively simple structures.
Nonetheless, this indicates the importance of providing clear guidance and ground truth exemplars to the crowd annotators by the researchers when setting up segmentation tasks.

\subsubsection{Crowd Segmentation in Different Modalities}
\label{result:Quality:Crowd Segmentation Versatility in Different Modalities}
It has been demonstrated that AI assistance in combination with instructions can dramatically improve the accuracy of labelling tasks.
To further investigate the versatility of our platform, we conducted tests involving AI-assisted labelling on MRI images from the MMWHS dataset, which differ significantly from the CT images from the same dataset, and CT images of the abdomen from the FLARE22 dataset. 

\begin{figure}[H]
    \vspace{-0.1cm}
    \centering
    \includegraphics[width=0.9\linewidth]{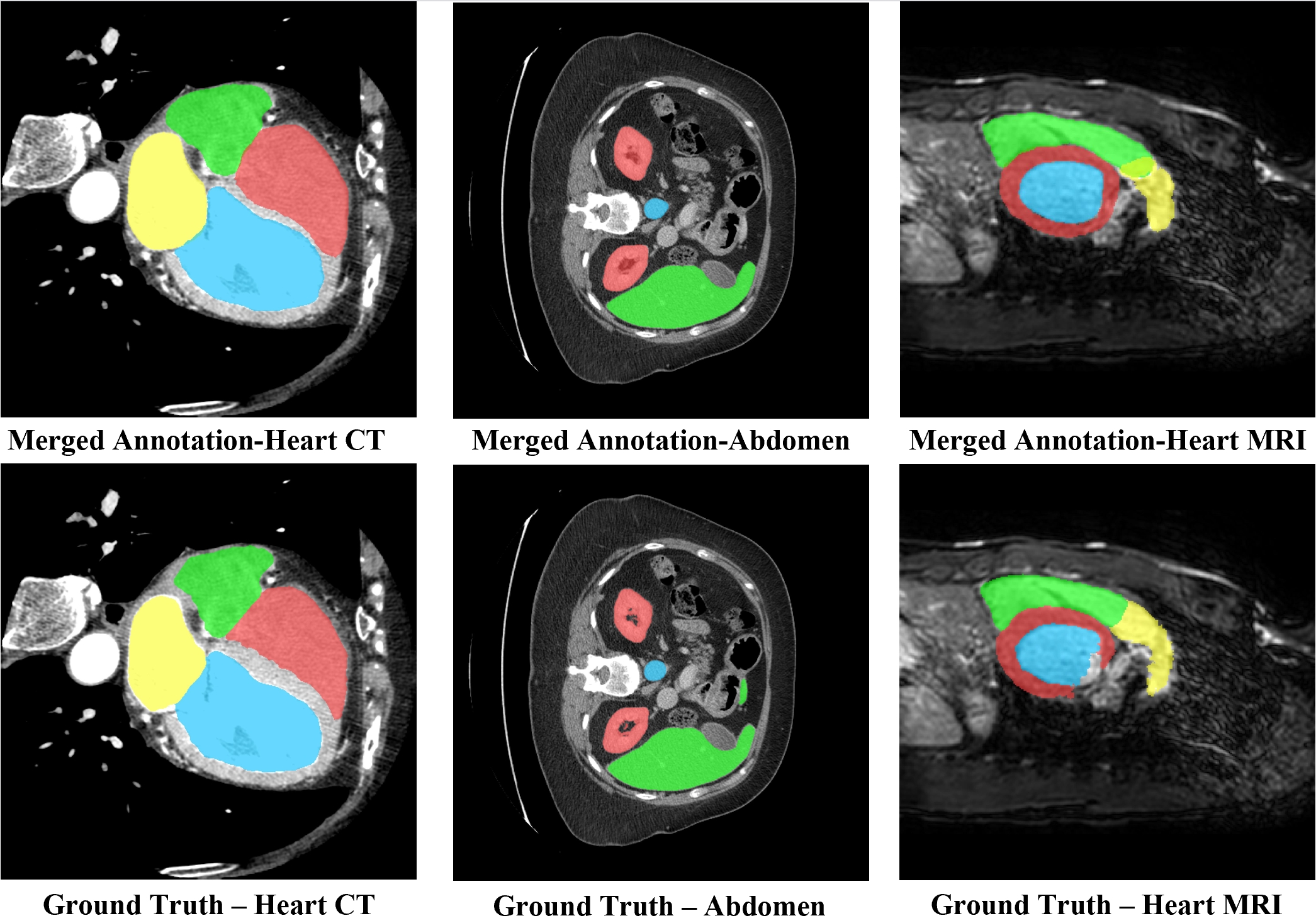}
    \vspace{-0.2cm}
    \caption{
        Comparison between the ground truth (Bottom) and merged crowd labels from different Datasets (Top): Task 3 MM-WHS CT (Left); Task 5 FLARE22 (Middle); and Task 6 MMWHS-MRI (Right)
    }
    \label{res:versatility}
\end{figure}

\vspace{-0.5cm}

It is notable that in Figure \ref{res:versatility}, visually the merged crowd segmentation is close to the ground truth, with the edges of the organ identified with high definition. 
Quantitatively, Table \ref{tab:T5FLARE22} shows that the labelling accuracy is very high in liver and kidney segmentation.
Specifically, the DSC is approximately 0.96 for both the liver and aorta (IoU of about 0.93) and about 0.75 for the kidney (IoU of about 0.65). 
According to Table \ref{tab:T6CT}, despite the complexity of MRI images, the DSC and IoU metrics are acceptable, yielding a DSC of about 0.7 (IoU of about 0.6) for all.
These results illustrate that our platform is versatile to ensure the accuracy of labelling tasks across different modalities of images. 
This endorses the customizability of the crowdsourcing platform by ensuring that different datasets can all be segmented efficiently by the merged crowd labels.

\subsection{Quantity -- Enlarging Dataset with Synthetic Data}
\label{result:Quantity}

\subsubsection{Synthetic Images in Different Modalities}
\label{result:Quantity:Versatility in Different Modalities}

\vspace{-0.4cm}
\begin{figure}[H]
    \centering
    \includegraphics[width=0.9\linewidth]{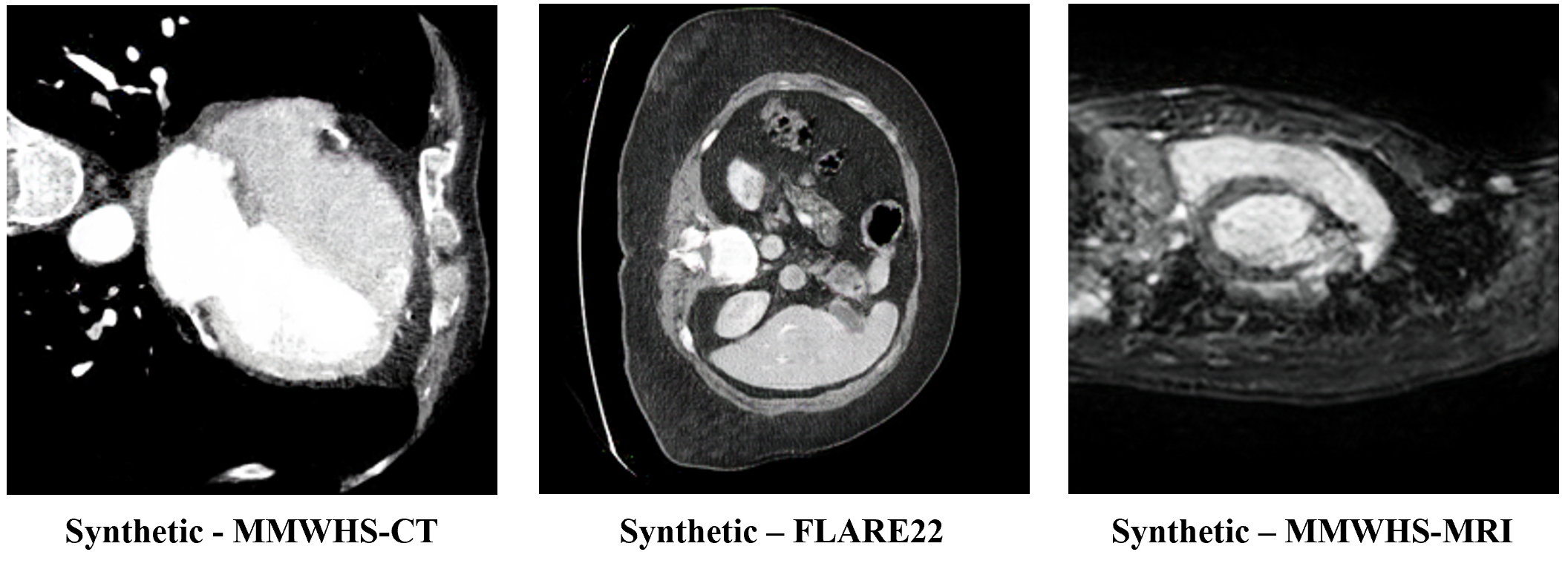}
    \caption{
        Exemplars of synthetic images generated by pix2pixGAN, using masks from: MMWHS-CT Slice 100 (Left); FLARE22 Slice 55 (Middle); MMWHS-MRI Slice 65 (Right)
    }
    \label{res:versatilityGAN}
\end{figure}
\vspace{-0.4cm}

To evaluate the versatility of the pix2pixGAN model, we trained it on datasets comprising different diverse segmentation pairs from medical imaging modalities and organs.
The results demonstrated the model's ability to generate synthetic images across different modalities, including MMWHS-CT, MMWHS-MRI, and FLARE22.
Furthermore, the model's versatility was evidenced by its capacity to generate clearly identifiable organs and compartment tissues, such as the heart in MMWHS, and the liver and kidneys in FLARE22, as we can see in Figure \ref{res:versatilityGAN}.
The generated synthetic images exhibited distinct edges and good contrast, particularly when multiple organs are present within a single image, with characterisable and identifiable morphology.
These findings demonstrated pix2pixGAN's high versatility in generating synthetic images across various modalities and anatomical structures in medical imaging.
However, it is noted that the synthetic image organs are often found at wrong vertebral levels, which indicates a lack of realism.
A potential improvement is suggested where landmarks apart from ROIs could be included during synthesis to refine anatomical accuracy.

\subsubsection{Efficiency of Enlarged Dataset}
\label{result:Quantity:Efficiency of Enlarged Dataset}

\begin{figure*}
    \begin{minipage}[b]{0.3\linewidth}
        \begin{table}[H]
            \centering
            \scalebox{0.7}{
                \begin{tabular}{ccccc}
                    \toprule[1.2pt]
                        & \multicolumn{2}{c}{Control} & \multicolumn{2}{c}{Enlarged} \\
                    ROI & DSC          & IoU          & DSC           & IoU          \\
                    \midrule[1.2pt]
                    LA  & 0.6318       & 0.6859       & 0.7767        & 0.7901       \\
                    RA  & 0.5992       & 0.6361       & 0.6988        & 0.7000       \\
                    LV  & 0.6025       & 0.5689       & 0.6821        & 0.6464       \\
                    RV  & 0.7254       & 0.7390       & 0.8086        & 0.7850       \\
                    \bottomrule[1.2pt]
                \end{tabular}
            }
            \vspace{-0.2cm}
            \caption{Mean DSC and IoU for UNet trained with control and enlarged MMWHS-CT dataset}
            \label{tab:CtrlVersa_MMWHS-CT}
        \end{table}
    \end{minipage}
    \hspace{0.025\linewidth}
    \begin{minipage}[b]{0.3\linewidth}
        \begin{table}[H]
            \centering
            \scalebox{0.7}{
                \begin{tabular}{ccccc}
                    \toprule[1.2pt]
                            & \multicolumn{2}{c}{Control}                       & \multicolumn{2}{c}{Enlarged}                      \\
                    ROI              & DSC          & IoU          & DSC           & IoU          \\
                    \midrule[1.2pt]
                    Liver  & 0.7511                  & 0.8695                  & 0.8092                  & 0.8369                  \\
                    Kidney & 0.5648                  & 0.7039                  & 0.6030                  & 0.8135                  \\
                    Aorta  & 0.1490                  & 0.6580                  & 0.3392                  & 0.7802                  \\
                    \bottomrule[1.2pt]
             \end{tabular}
            }
            \vspace{-0.2cm}
            \caption{Mean DSC and IoU for UNet trained with control and enlarged FLARE22 dataset}
            \label{tab:CtrlVersa_FLARE22}
        \end{table}
    \end{minipage}
    \hspace{0.025\linewidth}
    \begin{minipage}[b]{0.3\linewidth}
        \begin{table}[H]
            \centering
            \scalebox{0.7}{
                \begin{tabular}{ccccc}
                    \toprule[1.2pt]
                    & \multicolumn{2}{c}{Control} & \multicolumn{2}{c}{Enlarged} \\
                    ROI              & DSC          & IoU          & DSC           & IoU          \\
                    \midrule[1.2pt]
                    LV               & 0.6727       & 0.786        & 0.7689        & 0.7703       \\
                    Myocardium of LV & 0.7063       & 0.7281       & 0.7271        & 0.7734       \\
                    RV               & 0.7608       & 0.8606       & 0.8268        & 0.8591       \\
                    Pulmonary Artery & 0.4716       & 0.5629       & 0.5156        & 0.5513       \\
                    \bottomrule[1.2pt]
                \end{tabular}
            }
            \vspace{-0.2cm}
            \caption{Mean DSC and IoU for UNet trained with control and enlarged MMWHS-MRI dataset}
            \label{tab:CtrlVersa_MMWHS-MRI}
        \end{table}
    \end{minipage}
    \hfill
\end{figure*}

\begin{table*}
    \centering
    \scalebox{0.7}{
        \begin{tabular}{ccccccc}
            \toprule[1.2pt]
            & \multicolumn{2}{c}{Control}   & \multicolumn{2}{c}{Enlarged} & \multicolumn{2}{c}{Enhanced} \\
            ROI     & DSC           & IoU          & DSC           & IoU     & DSC           & IoU           \\
            \midrule[1.2pt]
            \multirow{2}{*}{Liver} & 0.7921                 & 0.9005   & 0.8546 & 0.9061        &0.8895 &0.9322       \\
                        & {[}0.7634 0.8209{]}    & {[}0.8928 0.9081{]} &[0.8377 0.8714]  &[0.8999 0.9123] &[0.8797 0.8993]  &[0.9271 0.9372]   \\
            \multirow{2}{*}{Aorta} & 0.1467                 & 0.4932    & 0.2604 & 0.7453       &0.5045       &0.7291      \\
                        & {[}0.1411 0.1522{]}    & {[}0.4659 0.5205{]}   &[0.2537 0.2671]  &[0.7161 0.7745] &[0.4928 0.5162]  &[0.7059 0.7523]   \\
            \bottomrule[1.2pt]
        \end{tabular}
    }
    \vspace{-0.2cm}
    \caption{Mean DSC and IoU with 95\% CI for UNet specifically trained for liver and aorta from control, enlarged, and enhanced FLARE22 dataset}
    \label{tab:EnhancedSpecific_FLARE22}
\end{table*}

To further evaluate the feasibility of using an enlarged dataset to improve model training in scenarios with limited data, we conducted a segmentation task comparison between the original and enlarged datasets.
For each of the MMWHS-CT, MMWHS-MRI, and FLARE22 datasets, 20 slices were randomly selected, with 10 used for training and 10 for testing.
Additionally, 10 synthetic images were generated from the ground truth labels of the training slices using pix2pixGAN.
For the control dataset, a UNet model was trained on 10 real images with their corresponding ground truth labels.
For the enlarged dataset, a UNet model was trained on 20 images, comprising of 10 real and 10 synthetic images, along with their ground truth labels.

Results from Table \ref{tab:CtrlVersa_MMWHS-CT}, \ref{tab:CtrlVersa_FLARE22}, and \ref{tab:CtrlVersa_MMWHS-MRI} suggest all modalities have shown notable improvements in training scores with the enlarged dataset, apart from few IoUs fluctuated at a slightly lower score.
Notably, Table \ref{tab:CtrlVersa_FLARE22} shows an average of 15.9\% increase for DSC and 11.1\% increase for IoU.
Aorta, as the hardest segmented ROI in FLARE22, has improved from a DSC value of 0.1490 (IoU of 0.6580) to a DSC value of 0.3392 (IoU of 0.7802).
Overall, this result validates the effectiveness of incorporating synthetic data to improve model training outcomes in data-limited scenarios.

\subsection{Enhanced Dataset -- Combining Generative AI and Crowdsourcing}
\label{result:temp}

Finally, we combined the high quality merged crowd labels with GAN enlarged dataset as an enhanced dataset to evaluate the potential to improve model training in limited data scenarios further.
To ensure the training dataset's quality, only 5 FLARE22 Liver and Aorta merged crowd labels are used for the enhanced dataset due to their high similarity to the ground truth, with DSC above 0.95 and IoU above 0.92 as shown in Table \ref{tab:T5FLARE22}.
Therefore, as a preliminary evaluation, we trained three segmentation UNet models for the Aorta and Liver using three versions of the FLARE22 dataset: a control dataset, an enlarged dataset, and an enhanced dataset.
The control dataset included 10 real images; the enlarged dataset consisted of 10 real images and 10 synthetic images; and the enhanced dataset comprised 10 real images, 10 synthetic images, and 5 merged crowd labels. 

The metrics present in Table \ref{tab:EnhancedSpecific_FLARE22} indicate a significant improvement ($p<0.001$ for both liver and aorta DSC using unpaired t-test) in segmentation accuracy from the control dataset to the enlarged dataset, with further enhancement ($p<0.001$ for both liver and aorta DSC using unpaired t-test) when utilising the enhanced dataset compared to the enlarged dataset.
Overall, the enhance dataset performance has significant improvement when compared with the control dataset ($p<0.0001$ for both liver and aorta DSC using unpaired t-test). 
Notably, the enhanced dataset achieves a 12.3\% increase in DSC for liver segmentation compared to the control dataset.
Furthermore, the DSC for the Aorta increase substantially, from 0.1467 to 0.5045, and IoU improve from 0.4932 to 0.7291, highlighting enhanced feature extraction for challenging segmented ROIs.

These findings validate the potential of augmenting a limited training dataset with GAN-generated synthetic images and high-quality merged crowd labels, supporting the feasibility of our proposed framework.

\section{Conclusion}
\label{sec:conclusion}
To conclude, it is evident that it is possible to improve the data-gathering rate to create a fully labelled dataset by using crowdsourcing. 
We demonstrated that using a flexible zero-shot segmentation AI model such as MedSAM can improve the user experience and efficiency of labelling. 
Including synthetic images generated by GAN models like pix2pixGAN to enlarge the dataset has been proven to improve the accuracy of the segmentation model. 
A prototype platform is implemented to demonstrate the workflow and can act as a provision for a more robust platform that can effectively collect labelling data from the crowd. 
Crowdsourcing can be included as a data-gathering pipeline for future researchers in training their AI models and algorithms. 
Building upon the foundation of this research, we demonstrated a framework exploiting the potential of coupling AI and crowdsourcing to resolve the scarcity in the availability of medical images for model training. 
Our framework is general and versatile, and can be extended by others to contribute and incorporate for specific modalities.

\section{Acknowledgements}

This study was supported in part by the ERC IMI (101005122), the H2020 (952172), the MRC (MC/PC/21013), the Royal Society (IEC/NSFC/211235),
the NVIDIA Academic Hardware Grant Program, the SABER project supported by Boehringer Ingelheim Ltd, NIHR Imperial Biomedical Research Centre (RDA01), Wellcome Leap Dynamic Resilience,
UKRI guarantee funding for Horizon Europe MSCA Postdoctoral Fellowships (EP/Z002206/1), and the UKRI Future Leaders Fellowship (MR/V023799/1).

\newpage
{\small
\bibliographystyle{ieee_fullname}
\bibliography{manuscript_final}

\begin{thebibliography}{10}\itemsep=-1pt

\bibitem{ali2022role}
Hazrat Ali, Md~Rafiul Biswas, Farida Mohsen, Uzair Shah, Asma Alamgir, Osama Mousa, and Zubair Shah.
\newblock The role of generative adversarial networks in brain mri: a scoping review.
\newblock {\em Insights into imaging}, 13(1):98, 2022.

\bibitem{alzubaidi2023survey}
Laith Alzubaidi, Jinshuai Bai, Aiman Al-Sabaawi, Jose Santamar{\'\i}a, Ahmed~Shihab Albahri, Bashar Sami~Nayyef Al-dabbagh, Mohammed~A Fadhel, Mohamed Manoufali, Jinglan Zhang, Ali~H Al-Timemy, et~al.
\newblock A survey on deep learning tools dealing with data scarcity: definitions, challenges, solutions, tips, and applications.
\newblock {\em Journal of Big Data}, 10(1):46, 2023.

\bibitem{creswell2018generative}
Antonia Creswell, Tom White, Vincent Dumoulin, Kai Arulkumaran, Biswa Sengupta, and Anil~A Bharath.
\newblock Generative adversarial networks: An overview.
\newblock {\em IEEE signal processing magazine}, 35(1):53--65, 2018.

\bibitem{HugginFaceE2024HugginFaceSpaces}
Huggin~Face Enterprise.
\newblock Huggin face spaces, 2024.
\newblock Available at: \url{https://huggingface.co/}; Accessed on 17th January 2024.

\bibitem{frid2018gan}
Maayan Frid-Adar, Idit Diamant, Eyal Klang, Michal Amitai, Jacob Goldberger, and Hayit Greenspan.
\newblock Gan-based synthetic medical image augmentation for increased cnn performance in liver lesion classification.
\newblock {\em Neurocomputing}, 321:321--331, 2018.

\bibitem{galbusera2024image}
Fabio Galbusera and Andrea Cina.
\newblock Image annotation and curation in radiology: an overview for machine learning practitioners.
\newblock {\em European Radiology Experimental}, 8(1):11, 2024.

\bibitem{goodfellow2014generative}
Ian Goodfellow, Jean Pouget-Abadie, Mehdi Mirza, Bing Xu, David Warde-Farley, Sherjil Ozair, Aaron Courville, and Yoshua Bengio.
\newblock Generative adversarial nets.
\newblock {\em Advances in neural information processing systems}, 27, 2014.

\bibitem{goodfellow2020generative}
Ian Goodfellow, Jean Pouget-Abadie, Mehdi Mirza, Bing Xu, David Warde-Farley, Sherjil Ozair, Aaron Courville, and Yoshua Bengio.
\newblock Generative adversarial networks.
\newblock {\em Communications of the ACM}, 63(11):139--144, 2020.

\bibitem{greenspan2016guest}
Hayit Greenspan, Bram Van~Ginneken, and Ronald~M Summers.
\newblock Guest editorial deep learning in medical imaging: Overview and future promise of an exciting new technique.
\newblock {\em IEEE transactions on medical imaging}, 35(5):1153--1159, 2016.

\bibitem{gurari2015collect}
Danna Gurari, Diane Theriault, Mehrnoosh Sameki, Brett Isenberg, Tuan~A Pham, Alberto Purwada, Patricia Solski, Matthew Walker, Chentian Zhang, Joyce~Y Wong, et~al.
\newblock How to collect segmentations for biomedical images? a benchmark evaluating the performance of experts, crowdsourced non-experts, and algorithms.
\newblock In {\em 2015 IEEE winter conference on applications of computer vision}, pages 1169--1176. IEEE, 2015.

\bibitem{heim2018large}
Eric Heim, Tobias Ro{\ss}, Alexander Seitel, Keno M{\"a}rz, Bram Stieltjes, Matthias Eisenmann, Johannes Lebert, Jasmin Metzger, Gregor Sommer, Alexander~W Sauter, et~al.
\newblock Large-scale medical image annotation with crowd-powered algorithms.
\newblock {\em Journal of Medical Imaging}, 5(3):034002--034002, 2018.

\bibitem{heimann2009statistical}
Tobias Heimann and Hans-Peter Meinzer.
\newblock Statistical shape models for 3d medical image segmentation: a review.
\newblock {\em Medical image analysis}, 13(4):543--563, 2009.

\bibitem{Amazon2024AWSMTurk}
Amazon Mechanical~Turk Inc.
\newblock Amazon mechanical turk (mturk), 2024.
\newblock Available at: \url{https://www.mturk.com/}; Accessed on 7th April 2024.

\bibitem{Duolingo2024Fraud}
Google Inc.
\newblock Duolingo fraud detection, 2024.
\newblock Available at: \url{https://www.duolingo.com/}; Accessed on 7th April 2024.

\bibitem{HumanSignal2024LabelStudio}
Human~Signal Inc.
\newblock Open source data labelling platform, 2023.
\newblock Available at: \url{https://labelstud.io/}; Accessed on 3rd November 2023.

\bibitem{NCICIP2022Cancer}
National~Cancer Institute.
\newblock Cip cancer imaging program, cancer imaging archive, 2015.
\newblock Available at: \url{https://www.cancerimagingarchive.net/}; Accessed on 27th December 2023.

\bibitem{isola2017image}
Phillip Isola, Jun-Yan Zhu, Tinghui Zhou, and Alexei~A Efros.
\newblock Image-to-image translation with conditional adversarial networks.
\newblock In {\em Proceedings of the IEEE conference on computer vision and pattern recognition}, pages 1125--1134, 2017.

\bibitem{kirillov2023segment}
Alexander Kirillov, Eric Mintun, Nikhila Ravi, Hanzi Mao, Chloe Rolland, Laura Gustafson, Tete Xiao, Spencer Whitehead, Alexander~C Berg, Wan-Yen Lo, et~al.
\newblock Segment anything.
\newblock In {\em Proceedings of the IEEE/CVF International Conference on Computer Vision}, pages 4015--4026, 2023.

\bibitem{Google2024reCAPTCHA}
Google LLC.
\newblock Google recaptcha, 2024.
\newblock Available at: \url{https://www.google.com/recaptcha/about/}; Accessed on 7th April 2024.

\bibitem{Appen2024Fig8}
Appen ltd.
\newblock Appen figure eight, 2024.
\newblock Available at: \url{https://www.appen.com/ai-data}; Accessed on 7th April 2024.

\bibitem{ma2022FLARE22}
Jun Ma.
\newblock Miccai flare22 challenge dataset (50 labeled abdomen ct scans), 2022.
\newblock Available at: \url{https://zenodo.org/records/7860267}; Accessed on 10th December 2023.

\bibitem{ma2024segment}
Jun Ma, Yuting He, Feifei Li, Lin Han, Chenyu You, and Bo Wang.
\newblock Segment anything in medical images.
\newblock {\em Nature Communications}, 15(1):654, 2024.

\bibitem{maier2014can}
Lena Maier-Hein, Sven Mersmann, Daniel Kondermann, Sebastian Bodenstedt, Alexandro Sanchez, Christian Stock, Hannes~Gotz Kenngott, Mathias Eisenmann, and Stefanie Speidel.
\newblock Can masses of non-experts train highly accurate image classifiers? a crowdsourcing approach to instrument segmentation in laparoscopic images.
\newblock In {\em Medical Image Computing and Computer-Assisted Intervention--MICCAI 2014: 17th International Conference, Boston, MA, USA, September 14-18, 2014, Proceedings, Part II 17}, pages 438--445. Springer, 2014.

\bibitem{mirza2014conditional}
Mehdi Mirza and Simon Osindero.
\newblock Conditional generative adversarial nets.
\newblock {\em arXiv preprint arXiv:1411.1784}, 2014.

\bibitem{mok2019learning}
Tony~CW Mok and Albert~CS Chung.
\newblock Learning data augmentation for brain tumor segmentation with coarse-to-fine generative adversarial networks.
\newblock In {\em Brainlesion: Glioma, Multiple Sclerosis, Stroke and Traumatic Brain Injuries: 4th International Workshop, BrainLes 2018, Held in Conjunction with MICCAI 2018, Granada, Spain, September 16, 2018, Revised Selected Papers, Part I 4}, pages 70--80. Springer, 2019.

\bibitem{musalamadugu2023generative}
Tanmai~Sree Musalamadugu and Hemachandran Kannan.
\newblock Generative ai for medical imaging analysis and applications.
\newblock {\em Future Medicine AI}, 1(0):FMAI5, 2023.

\bibitem{petrovic2020crowdsourcing}
Nata{\v{s}}a Petrovi{\'c}, Gabriel Moy{\`a}-Alcover, Javier Varona, and Antoni Jaume-i Cap{\'o}.
\newblock Crowdsourcing human-based computation for medical image analysis: A systematic literature review.
\newblock {\em Health Informatics Journal}, 26(4):2446--2469, 2020.

\bibitem{raykar2010learning}
Vikas~C Raykar, Shipeng Yu, Linda~H Zhao, Gerardo~Hermosillo Valadez, Charles Florin, Luca Bogoni, and Linda Moy.
\newblock Learning from crowds.
\newblock {\em Journal of machine learning research}, 11(4), 2010.

\bibitem{sarker2021deep}
Iqbal~H Sarker.
\newblock Deep learning: a comprehensive overview on techniques, taxonomy, applications and research directions.
\newblock {\em SN Computer Science}, 2(6):420, 2021.

\bibitem{Shorten2019}
Connor Shorten and Taghi~M. Khoshgoftaar.
\newblock A survey on image data augmentation for deep learning.
\newblock {\em Journal of Big Data}, 6(1):60, Jul 2019.

\bibitem{sorensen1948method}
Thorvald Sorensen.
\newblock A method of establishing groups of equal amplitude in plant sociology based on similarity of species content and its application to analyses of the vegetation on danish commons.
\newblock {\em Biologiske skrifter}, 5:1--34, 1948.

\bibitem{walsh2020imaging}
Simon~LF Walsh, Stephen~M Humphries, Athol~U Wells, and Kevin~K Brown.
\newblock Imaging research in fibrotic lung disease; applying deep learning to unsolved problems.
\newblock {\em The Lancet Respiratory Medicine}, 8(11):1144--1153, 2020.

\bibitem{woznicki2023addressing}
Piotr Woznicki, Fabian~Christopher Laqua, Adam Al-Haj, Thorsten Bley, and Bettina Bae{\ss}ler.
\newblock Addressing challenges in radiomics research: systematic review and repository of open-access cancer imaging datasets.
\newblock {\em Insights into Imaging}, 14(1):216, 2023.

\bibitem{xing2024aieatsitselfcaveats}
Xiaodan Xing, Fadong Shi, Jiahao Huang, Yinzhe Wu, Yang Nan, Sheng Zhang, Yingying Fang, Mike Roberts, Carola-Bibiane Schönlieb, Javier~Del Ser, and Guang Yang.
\newblock When ai eats itself: On the caveats of data pollution in the era of generative ai, 2024.

\bibitem{zhuang2022MMWHS}
Xiahai Zhuang.
\newblock Mm-whs: Multi-modality whole heart segmentation, 2019.
\newblock Available at: \url{https://zmiclab.github.io/zxh/0/mmwhs/}; Accessed on 14th September 2023.

\bibitem{zhuang2019evaluation}
Xiahai Zhuang, Lei Li, Christian Payer, Darko {\v{S}}tern, Martin Urschler, Mattias~P Heinrich, Julien Oster, Chunliang Wang, {\"O}rjan Smedby, Cheng Bian, et~al.
\newblock Evaluation of algorithms for multi-modality whole heart segmentation: an open-access grand challenge.
\newblock {\em Medical image analysis}, 58:101537, 2019.

\end{thebibliography}
}

\end{document}